\documentclass[10pt, conference, compsocconf]{IEEEtran}
\IEEEoverridecommandlockouts
\ifCLASSINFOpdf
  \usepackage[pdftex]{graphicx}
  \graphicspath{{../pdf/}{../jpeg/}}
  \DeclareGraphicsExtensions{.pdf,.jpeg,.png}
\else
\fi
\usepackage[cmex10]{amsmath}
\usepackage{algpseudocode}
\algtext*{EndWhile}
\algtext*{EndIf}
\algtext*{EndFor}
\algtext*{EndFunction}
\algtext*{EndProcedure}
\usepackage{url}

\usepackage{tikz}
\usepackage{pgfplots}

\begin{document}
%
\title{An Easy-to-use Scalable Framework for Parallel Recursive Backtracking}




%
\author{
\IEEEauthorblockN{
Faisal N. Abu-Khzam\IEEEauthorrefmark{4},
Khuzaima Daudjee\IEEEauthorrefmark{2}$^{\star}$,
Amer E. Mouawad\IEEEauthorrefmark{2}$^{\star}$ and
Naomi Nishimura\IEEEauthorrefmark{2}$^{\star}$
\thanks{$^{\star}$ Research supported by the Natural Science and Engineering Research Council of Canada.} \\\\}
\IEEEauthorblockA{\IEEEauthorrefmark{4}Department of Computer Science and Mathematics\\
Lebanese American University\\
Beirut, Lebanon\\ Email: faisal.abukhzam@lau.edu.lb\\\\}
\IEEEauthorblockA{\IEEEauthorrefmark{2}David R. Cheriton School of Computer Science\\
University of Waterloo\\
Waterloo, Ontario, N2L 3G1, Canada\\ Email: \{kdaudjee, aabdomou, nishi\}@uwaterloo.ca}
}


\maketitle

\begin{abstract}
Supercomputers are equipped with an increasingly large number of cores
to use computational power as a way of solving problems that are otherwise intractable.
Unfortunately, getting serial algorithms to run in parallel to take advantage of these
computational resources remains a challenge for several application domains.
Many parallel algorithms can scale to only hundreds of cores. The limiting factors
of such algorithms are usually communication overhead and poor load balancing.
Solving NP-hard graph problems to optimality using exact algorithms
is an example of an area in which there has so far been
limited success in obtaining large scale parallelism.
Many of these algorithms use recursive backtracking as their core solution paradigm.
In this paper, we propose a lightweight, easy-to-use, scalable framework for transforming
almost any recursive backtracking algorithm into a parallel one.
Our framework incurs minimal communication overhead and guarantees a
load-balancing strategy that is implicit, i.e., does not require any problem-specific knowledge.
The key idea behind this framework is the use of an
indexed search tree approach that is oblivious to the problem being solved.
We test our framework with parallel implementations of algorithms for the well-known
Vertex Cover and Dominating Set problems.
On sufficiently hard instances, experimental results show linear speedups
for thousands of cores, reducing running times from days to just a few minutes.
\end{abstract}


%
\IEEEpeerreviewmaketitle

\section{Introduction}
Parallel computation is becoming increasingly important as performance levels out
in terms of delivering parallelism within a single processor due to Moore's law.
This paradigm shift means that to attain speed-up, software that
implements algorithms that can run in
parallel on multiple cores is required.  Today we have a
growing list of supercomputers with tremendous processing power. Some
of these systems include more than a million computing cores
and can achieve up to 30 Petaflop/s.  The constant increase in the
number of cores per supercomputer motivates the development of
parallel algorithms that can efficiently utilize such processing
infrastructures. Unfortunately, migrating known serial algorithms to
exploit parallelism while maintaining scalability is not a
straightforward job.  The overheads introduced by parallelism are very
often hard to evaluate, and fair load balancing is possible only when
accurate estimates of task ``hardness'' or ``weight'' can be
calculated on-the-fly.  Providing such estimates usually requires
problem-specific knowledge, rendering the techniques developed for
problem $A$ useless when trying to parallelize an algorithm for problem $B$.

As it is not likely that polynomial-time algorithms can be found for
NP-hard problems, the search for fast deterministic algorithms
could benefit greatly from the processing capabilities of
supercomputers.  Researchers working in the area of exact algorithms
have developed algorithms yielding lower and lower running
times~\cite{CKJ01, FKW05, CKX10, FGK05, RND09}.
However the major focus has been on improving
the worst-case behavior of algorithms (i.e. improving the best known
asymptotic bound on the running time of algorithms).  The practical
aspects of the possibility of exploiting parallel infrastructures has
received much less attention.

Most existing exact algorithms for NP-hard graph problems follow the
well-known branch-and-reduce paradigm.  A branch-and-reduce algorithm
searches the complete solution space of a given problem for an optimal
solution.  Simple enumeration is normally prohibitively expensive due
to the exponentially increasing number of potential solutions.  To
prune parts of the solution space, an algorithm uses reduction rules
derived from bounds on the function to be optimized and the value of
the current best solution.  The reader is referred to Woeginger's
excellent survey paper on exact algorithms for further
detail~\cite{W03}.  At the implementation level, branch-and-reduce
algorithms translate to search-tree-based recursive backtracking
algorithms. The search tree size usually grows exponentially with
either the size of the input instance $n$ or, when the problem is
fixed-parameter tractable, some integer parameter $k$ \cite{DF97}.
Nevertheless, search trees are good candidates for parallel
decomposition. Given $c$ cores, a brute-force parallel solution would
divide a search tree into $c$ subtrees and assign each subtree to a
separate core for sequential processing. One might hope to thus reduce
the overall running time by a factor of $c$.  However, this intuitive
approach suffers from several drawbacks, including the obvious lack of
load balancing.

Even though our focus is on NP-hard graph problems, we note
that recursive backtracking is a
widely-used technique for solving a very long list of practical
problems. This justifies the need for a general framework to simplify
the migration from serial to parallel.  One example of a successful
parallel framework for solving a different kind of problem is
MapReduce \cite{DG08}. The success of the MapReduce model can be
attributed to its simplicity, transparency, and scalability,
all of which are properties essential for any efficient parallel framework.  In this
paper, we propose a simple, lightweight, scalable framework for
transforming almost any recursive backtracking algorithm into a
parallel one with minimal communication overhead and a load balancing strategy
that is implicit, i.e., does not require any problem-specific knowledge.
The key idea behind the framework is the use of an indexed search-tree
approach that is oblivious to the problem being solved. To test our
framework, we use it to implement parallel exact algorithms for the
well-known {\sc Vertex Cover} and {\sc Dominating Set} problems.
Experimental results show that, on sufficiently hard instances, we
obtain linear speedups for at least 32,768 cores.

\section{Preliminaries}
Typically, a recursive backtracking algorithm exhaustively explores a search tree $T$
using depth-first search traversal and backtracking at the leaves of $T$.
Each node of $T$ (a {\em search-node})
can be denoted by $N_{d,p}$ for $d$
the depth of $N_{d,p}$ in $T$ and $p$ the position
of $N_{d,p}$ in the left-to-right ordering of all search-nodes at depth $d$.
The root of $T$ is thus $N_{0,0}$. We use $T(N_{d,p})$ to denote the subtree rooted at node $N_{d,p}$.
We say $T$ has {\em branching factor} $b$ if every search-node has at most $b$ children.
An example of a generic serial recursive backtracking
algorithm, {\sc Serial-RB}, is given in Figure \ref{serial-rb}.
The goal of this paper is to provide a framework to transform {\sc Serial-RB}
into an efficient parallel algorithm with as little effort as possible.
For ease of presentation, we make the following assumptions:

\begin{itemize}
\item[-] {\sc Serial-RB} is solving an NP-hard optimization problem (i.e. minimization or maximization) where each solution appears in a leaf of the search tree.
\item[-] The global variable $best\_so\_far$ stores the best solution found so far.
\item[-] The {\sc IsSolution($N_{d,p}$)} function returns true only if $N_{d,p}$ is a solution
which is ``better'' than $best\_so\_far$.
\item[-] The search tree explored by {\sc Serial-RB} is binary (i.e. every search-node has at most two children).
\end{itemize}

\noindent
None of these assumptions are needed to apply our framework. In Section \ref{sec-not-binary}, we
discuss how the same techniques can be easily adapted to any search tree with arbitrary branching factor.
The only (minor) requirement we impose
is that the number of children of a search-node can be calculated on-the-fly and
that generating those children (using  {\sc GetNextChild($N_{d,p}$)})
follows a deterministic procedure with a well-defined order.
In other words, if we run {\sc Serial-RB} an arbitrary
number of times on the same input instance, the search trees of all executions
will be identical. The reason for this restriction will become obvious later.

\begin{figure}[!t]
\centering
\begin{algorithmic}[1]
\Procedure{Serial-RB}{$N_{d,p}$}
\If{(\Call{IsSolution}{$N_{d,p}$})}
  \State $best\_so\_far \gets N_{d,p}$;
\EndIf
\If{(\Call{IsLeaf}{$N_{d,p}$})}
  \State Apply backtracking;\Comment{undo operations}
\EndIf
\State $p' \gets 0$;
\While{\Call{HasNextChild}{$N_{d,p}$}}
    \State $N_{d + 1, p'} \gets \Call{GetNextChild}{N_{d,p}}$;
    \State \Call{Serial-RB}{$N_{d + 1, p'}$};
    \State $p' \gets p' + 1$;
\EndWhile
\EndProcedure
\end{algorithmic}
\caption{The {\sc Serial-RB} algorithm}
\label{serial-rb}
\end{figure}

In a parallel environment, we denote by ${\cal C} = \{C_0, C_1,
\ldots, C_c\}$ the set of available computing cores. The {\em
  rank} of $C_i$ is equal to $i$ and $|{\cal C}| = c$.  We use the
terms {\em worker} and {\em core}
interchangeably to refer to some $C_i$ participating in a parallel
computation.  Each search-node in $T$ corresponds to a {\em task}, where
tasks are {\em exchanged} between cores using some specified encoding.  We
use $E(N_{d,p})$ to denote the encoding of $N_{d,p}$.  When
search-node $N_{d,p}$ is assigned to $C_i$, we say $N_{d,p}$ is the
{\em main task} of $C_i$. Any task in $T(N_{d,p})$ sent from $C_i$ to
some $C_j$, $i \neq j$, is a {\em subtask} for $C_i$ and becomes the
main task of $C_j$.
The {\em weight of a task}, $w(N_{d,p})$, is a
numerical value indicating the estimated completion time for $N_{d,p}$
relative to other tasks.  That is, when $w(N_{d,p}) > w(N_{d',p'})$,
we expect the exploration of $T(N_{d,p})$ to require more
computational time than $T(N_{d',p'})$.  Task weight plays a crucial
role in the design of efficient dynamic load-balancing strategies
\cite{DB07,WERL11,AM12}. Without any problem-specific knowledge, the ``best''
indicator of the weight of $N_{d,p}$ is nothing but $d$ since estimating
the size of $T(N_{d,p})$ is almost impossible.  We capture this
notion by setting $w(N_{d,p}) = {1 \over {d + 1}}$.  We say a task is
{\em heavy} if it has weight close to $1$ and {\em light} otherwise.

From the standpoint of high-performance computing, practical parallel
exact algorithms for hard problems mean one thing: unbounded scalability.
The seemingly straight-forward parallel nature of
search-tree decomposition is deceiving: previous
work has shown that attaining scalability
is far from easy \cite{K88,SZJK11,RLS04}.
To the best of our knowledge, the most efficient existing parallel algorithms
that solve problems similar to those we consider
were only able to scale to less than a few
thousand (or only a few hundred) cores \cite{WERL11,AM12,ARAS08}.
One of our main motivations was to solve
extremely hard instances of the {\sc Vertex Cover} problem such as
the 60-cell graph \cite{DDMSWWFDD13}.
In earlier work, we first attempted to tackle the problem by improving
the efficiency of our serial algorithm \cite{ALMN10}. Alas, some instances remained unsolved
and some required several days of execution before we could obtain a solution.
The next obvious step was to attempt a parallel implementation. As we encountered
scalability issues, it became clear that solving such instances in an ``acceptable'' amount of
time would require a scalable algorithm that can effectively
utilize much more than the 1,024-core limit we attained in previous work~\cite{AM12}.

We discuss the lessons we have learned and what we believe to be the main
reasons of such poor scalability in Section \ref{related-work}.
In Section \ref{framework}, we present the main concepts and
strategies we use to address these challenges under our parallel
framework. Finally, implementation details and experimental results are covered in
Sections \ref{implementation} and \ref{exp-results}, respectively.

\section{Challenges and Related Work}\label{related-work}

\subsection{Communication Overhead}
The most evident overhead in parallel algorithms is that of communication.
Several models have already been presented in the literature
including centralized (i.e. the master-worker(s) model where most of
the communication and task distribution duties
are assigned to a single core) \cite{ARAS08},
decentralized \cite{DB07, AM12}, or a hybrid of both \cite{SZJK11}.
Although each model has its pros and cons,  we believe
that centralization rapidly becomes a bottleneck when the
number of computing cores exceeds a certain threshold.
Even though our framework can be implemented
under any communication model, we chose
to follow a simplified version of our previous decentralized
approach~\cite{AM12}.

An efficient communication model has to (i) reduce the total number of message
transmissions and (ii) minimize the travel distance (number of hops)
for each transmission. Unfortunately, (ii) requires detailed
knowledge of the underlying network
architecture and comes at the cost of portability.
For (i), the message complexity is tightly coupled
with the number of times each $C_i$
runs out of work and requests more. Therefore, to minimize the
number of generated messages, we need to
maximize ``work time'', which is achieved by better dynamic load balancing.

\subsection{Tasks, Buffers, and Memory Overhead}\label{sec-buffers}
No matter what communication model is used, a certain encoding
has to be selected for representing tasks in memory.
An obvious drawback of the encoding used by Finkel and Manber~\cite{FM87} is that
every task is an exact copy of a search-node, whose size can be quite large.
In a graph algorithm, every search-node might contain a modified
version of the input graph (and some additional information).
In this case, a more compact task-encoding scheme
is needed to reduce both the memory and communication overheads.

Almost all parallel algorithms in the literature require a
task-buffer or task-queue to store multiple tasks for eventual
delegation \cite{AM12, SZJK11, ARAS08, FM87}.  As buffers have limited
size, their usage requires the selection of a ``good'' parameter value
for task-buffer size.  Choosing the size can be a daunting task, as
this parameter introduces a tradeoff between the amount of time spent
on creating or sending tasks and that spent on solving tasks.  It is
very common for such parallel algorithms to enter a loop of multiple
light task exchanges which unnecessarily consumes considerable amounts
of time and memory \cite{AM12}.  Tasks in such loops would have been
more efficiently solved in-place by a single core.

\subsection{Initial Distribution}

Efficient dynamic load balancing is key to scalable parallel
algorithms.  To avoid loops of multiple light task exchanges, initial
task distribution also plays a major role.  Even with clever
load-balancing techniques, such loops can consume a lot of resources
and delay (or even deny) the system from reaching a balanced state.

\subsection{Serial Overhead}
All the items discussed above induce some serial overhead. Here we
focus on encoding and decoding of tasks, which greatly affect the
performance of any parallel algorithm.  Upon receiving a new task,
each computing core has to perform a number of operations to correctly
restart the search-phase, i.e. resume the exploration of its assigned subtree.
When the search reaches the bottom levels of the tree, the amount
of time required to start a task might exceed the time required to solve it,
a situation that should be avoided.  Encoding tasks and storing them
in buffers also consumes time. In fact, the more we attempt to
compress task encodings the more serial work is required for decoding.

For NP-hard problems, it is important to account for what we call the {\em butterfly effect}
of polynomial overhead. Since the size of the search tree is usually exponential in the size of the
input, any polynomial-time (or even constant-time) operations can have significant effects
on the overall running times \cite{ALMN10}, by virtue of being executed exponentially many times.
In general, the disruption time (time spent doing
non-search-related work) has to be minimized.

\subsection{Load Balancing}
Task creation is, we believe, the most critical factor affecting load
balancing.  Careful tracing of recursive backtracking algorithms shows
that most computational time is spent in the bottom of the search
tree, where $d$ is very large.  Moreover, since task-buffers have
fixed size, any parallel execution of a recursive backtracking
algorithm relying on task-buffers is very likely to reach a state
where all buffers contain light tasks.  Loops of multiple light task
exchanges most often occur in such scenarios.  To avoid them, we need a
mechanism that enables the extraction of a task of maximum weight from
the subtree assigned to a $C_i$, that is, the highest unvisited node
in the subtree assigned to $C_i$.

Several load-balancing strategies have been proposed
in the literature \cite{K88,KGV94}. In recent work \cite{WERL11}, a load-balancing
strategy designed specifically for the {\sc Vertex Cover} problem was presented.
The algorithm is based on a dynamic master-worker model
where prior knowledge about generated instances is manipulated
so that the core having the estimated heaviest
task is selected as master. However, scalability of this
approach was limited to only 2,048 cores.

\subsection{Termination Detection}
In a centralized model, the master detects termination using
straightforward protocols.  The termination protocol can be initiated
several times by different cores in a decentralized environment,
rendering detection more challenging. In this work, we use a protocol
similar to the one proposed in \cite{AM12}, where each core, which can be in
one of three states, broadcasts any state change to all other cores.

\subsection{Problem Independence}
We need to address all the challenges listed above independently of the problem being solved.
Moreover, we want to minimize the amount of work required for migrating existing
serial algorithms to parallel ones.

\section{The Framework}\label{framework}
In this section, we present our framework by showing how
to incrementally transform {\sc Serial-RB} into
a parallel algorithm. First, we discuss indexed search trees and their
use in a generic and compact task-encoding scheme.
As a byproduct of this encoding, we show how we can efficiently guarantee the extraction
of the heaviest unprocessed task for dynamic load balancing.
We provide pseudocode to illustrate the simplicity
of transforming serial algorithms to parallel ones.
The end result is a parallel algorithm, {\sc Parallel-RB}, which
consists of two main procedures: {\sc Parallel-RB-Iterator} and  {\sc Parallel-RB-Solver}.

\subsection{Indexed Search Trees}
For a binary search tree $T$, we let $left(N_{d,p})$ and $right(N_{d,p})$ denote the left child and the right child
of node $N_{d,p}$, respectively. We use the following procedure to assign an index, $idx$,
to every search-node in $T$ (where $\mid\mid$ denotes concatenation):

\begin{itemize}
\item[(1)] The root of $T$ has index 1 ($idx(N_{0,0}) = 1$)
\item[(2)] For any node $N_{d,p}$ in $T$:
\begin{itemize}
\item[] $idx(left(N_{d,p})) = idx(N_{d,p}) \mid\mid 0$ and
\item[] $idx(right(N_{d,p})) = idx(N_{d,p}) \mid\mid 1$
\end{itemize}
\end{itemize}

\begin{figure}[!t]
\centering
\begin{tikzpicture}[
level 1/.style={sibling distance = 4cm, level distance = 1cm},
level 2/.style={sibling distance = 2cm},
level 3/.style={sibling distance = 1cm},
label distance=-1mm,
every node/.style={circle, draw=black,thin, minimum size = 0.5cm},
emph/.style={edge from parent/.style={red,very thick,draw}},
norm/.style={edge from parent/.style={black,thin,draw}}
]
\node [label={[label distance=-1.5mm]450:{\small{$1$}}},scale=0.8]{$N_{0,0}$}
  child {node[label={[label distance=-2mm]450:{\small{$10$}}},scale=0.8] {$N_{1,0}$}
    child {node[label={[label distance=-2mm]450:{\small{$100$}}},scale=0.8] {$N_{2,0}$}
        child {node[label={[label distance=-3mm]270:{\small{$1000$}}},scale=0.8] {$N_{3,0}$}}
        child {node[label={[label distance=-3mm]270:{\small{$1001$}}},scale=0.8] {$N_{3,1}$}}
    }
    child {node[label={[label distance=-2mm]450:{\small{$101$}}},scale=0.8] {$N_{2,1}$}
        child {node[label={[label distance=-3mm]270:{\small{$1010$}}},scale=0.8] {$N_{3,2}$}}
        child {node[label={[label distance=-3mm]270:{\small{$1011$}}},scale=0.8] {$N_{3,3}$}}
    }
  }
  child {node[label={[label distance=-2mm]450:{\small{$11$}}},scale=0.8] {$N_{1,1}$}
    child {node[label={[label distance=-2mm]450:{\small{$110$}}},scale=0.7] {$N_{2,2}$}
        child {node[label={[label distance=-3mm]270:{\small{$1100$}}},scale=0.8] {$N_{3,4}$}}
        child {node[label={[label distance=-3mm]270:{\small{$1101$}}},scale=0.8] {$N_{3,5}$}}
    }
    child {node[label={[label distance=-2mm]450:{\small{$111$}}},scale=0.8] {$N_{2,3}$}
        child {node[label={[label distance=-3mm]270:{\small{$1110$}}},scale=0.8] {$N_{3,6}$}}
        child {node[label={[label distance=-3mm]270:{\small{$1111$}}},scale=0.8] {$N_{3,7}$}}
    }
  };
\end{tikzpicture}
\caption{Example of an indexed binary search tree}
\label{fig-tree}
\end{figure}
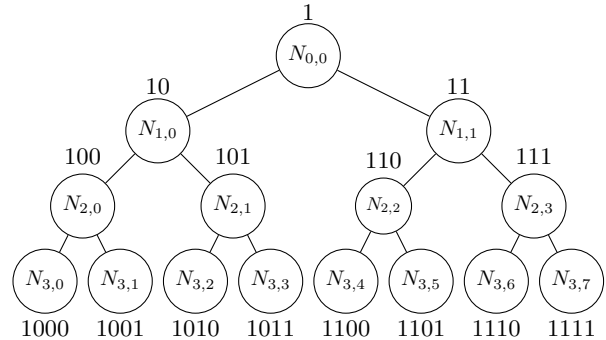

\noindent An example of an indexed binary search tree is given in Figure \ref{fig-tree}.
Note that this indexing method can easily be extended for arbitrary branching factor by
simply setting the index of the $k^{th}$ child of $N_{d,p}$ to $idx(N_{d,p}) \mid\mid (k - 1)$.

To incorporate the notion of indices, we introduce minor modifications to {\sc Serial-RB}.
We call this new version {\sc Almost-Parallel-RB} (Figure \ref{almost-parallel-rb}).
{\sc Almost-Parallel-RB} includes a global integer array
$current\_idx$ that is maintained by a single statement:
$current\_idx[d] = p$.  We let $E(N_{d,p}) = idx(N_{d,p})$; the
encoding of a task under our parallel framework corresponds to its
index and is of ${\mathcal{O}}(d)$ size.  Combined with an effective
load-balancing strategy which generates tasks having only small $d$
(i.e. heavy tasks), this approach greatly reduces memory and
communication overhead. Upon receiving an encoded task, every core now
requires an additional user-provided function {\sc ConvertIndex}, the
implementation details of which are problem-specific (Section
\ref{implementation} discusses some examples).  The purpose of this
function is to convert an index into an actual task from which the
search can proceed.  Since every index encodes the unique path from the root
of the tree to the corresponding search-node and by assumption
search-nodes are generated in a well-defined order, to retrace the
operations it suffices to iterate over the index.  Note that
the overhead introduced by this approach is closely related to the
number of tasks solved by each core.  Minimizing this number also
minimizes disruption time since the search-phase of the algorithm is
not affected.

We use the functions {\sc GetHeaviestTaskIndex} and {\sc FixIndex} (Figure~\ref{index-functions}) to repeatedly
extract the heaviest task from the $current\_idx$ array while ensuring
that no search-node is ever explored twice.
The general idea of indexing is not new and has been previously used for
prioritizing tasks in buffers or queues \cite{SZJK11}.
However, our approach completely eliminates the need for task-buffers,
effectively reducing the memory footprint of our algorithms
and eliminating the burden of selecting appropriate size parameters for
each buffer or task granularity as defined in \cite{SZJK11}.

In a parallel computation involving cores $C_i$ and $C_j$ in
Figure~\ref{fig-tree}, $C_i$ has main task $N_{0,0}$ and is currently
exploring node $N_{3,2}$ (hence $current\_idx = \{1,0,1,0\}$).  After
receiving an initial task request from $C_j$, $C_i$ calls {\sc
  GetHeaviestTaskIndex}, which returns $temp\_idx = \{1,-1\}$ and sets
$current\_idx = \{1,-1,1,0\}$.
(We use $current\_idx[l; h]$ to denote a subarray of $current\_idx$
starting at position $l$ and ending at position $h$.)
At the receiving end, $C_j$ calls {\sc
  FixIndex}, after which $temp\_idx = \{1,1\}$.  As seen in
Figure~\ref{fig-tree}, $N_{1,1}$ was in fact the heaviest
task in $T(N_{0,0})$.  If $C_j$ subsequently requests a second task from $C_i$
while $C_i$ is still working
on node $N_{3,2}$, the resulting task is $\{1,0,1,1\}$ and the
$current\_idx$ of $C_i$ is updated to $\{1,-1,1,-1\}$.

\begin{figure}[!t]
\centering
\begin{algorithmic}[1]
\Procedure{Almost-Parallel-RB}{$N_{d,p}$}
\If{($current\_idx[d] = -1$)}
  \State \textbf{terminate};
\EndIf
\State $current\_idx[d] \gets p$;
\If{(\Call{IsSolution}{$N_{d,p}$})}
  \State $best\_so\_far \gets N_{d,p}$;
\EndIf
\If{(\Call{IsLeaf}{$N_{d,p}$})}
  \State Apply backtracking;\Comment{undo operations}
\EndIf
\If{(\Call{TaskRequestExists}())}
    \State $x \leftarrow \Call{GetHeaviestTaskIndex}{current\_idx}$;
    \State \Call{send}{$x$, $requester$};
\EndIf
\State $p' \gets 0$;
\While{\Call{HasNextChild}{$N_{d,p}$}}
    \State $N_{d + 1, p'} \gets \Call{GetNextChild}{N_{d,p}}$;
    \State \Call{Almost-Parallel-RB}{$N_{d + 1, p'}$};
    \State $p' \gets p' + 1$;
\EndWhile
\EndProcedure
\end{algorithmic}
\caption{The {\sc Almost-Parallel-RB} algorithm}
\label{almost-parallel-rb}
\end{figure}

\begin{figure}[!t]
\centering
\begin{algorithmic}[1]
\Function{GetHeaviestTaskIndex}{$current\_idx$}
\For{$i \gets 0$, $current\_idx.length - 1$}
    \If{$(current\_idx[i] = 0)$}
      \State $current\_idx[i] \gets -1$;
      \State $temp\_idx \gets current\_idx[0; i]$;\Comment{subarray}
      \State \textbf{return} $temp\_idx$;
    \EndIf
\EndFor
\State \textbf{return} null;
\EndFunction
\end{algorithmic}
\begin{algorithmic}[1]
\Function{FixIndex}{$temp\_idx$}
\For{$i \gets 0$, $temp\_idx.length - 2$}
    \If{$(temp\_idx[i] < 0)$}
      \State $temp\_idx[i] \gets 0$;
    \EndIf
\EndFor
\State $temp\_idx[temp\_idx.length - 1] \gets 1$;
\State \textbf{return} $temp\_idx$;
\EndFunction
\end{algorithmic}
\caption{The {\sc GetHeaviestTaskIndex} and {\sc FixIndex} functions}
\label{index-functions}
\end{figure}

\noindent
Before exploring a search-node, every core must first use
$current\_idx$ to validate that the current branch was not previously
delegated to a different core (Figure \ref{almost-parallel-rb}, lines
2--3).  Whenever $C_i$ discovers a negative value in
$current\_idx[d]$, the search can terminate, since the remaining
subtree has been reassigned to a different core.

\subsection{From Serial to Parallel}
The {\sc Almost-Parallel-RB} algorithm is lacking a formal definition of a communication model
as well as the implementation details of the initialization and termination protocols.
For the former, we use a simple decentralized strategy in which any two cores can communicate.
We assume each core is assigned a unique rank $r$, for $0 \leq r < c$.
There are three different types of message exchanges under our framework: status updates, task requests or responses,
and notification messages. Each core can be in one of three states: active, inactive, or dead.
Before changing states, each core must broadcast a status update message to all participants.
This information is maintained by each core in a global integer array $statuses$.
Notification messages are optional broadcast messages whose purpose is to inform the
remaining participants of current progress. In our implementation, notification messages
are sent whenever a new solution is found. The message includes the size of the new solution which,
for many algorithms, can be used as a basis for effective pruning rules.

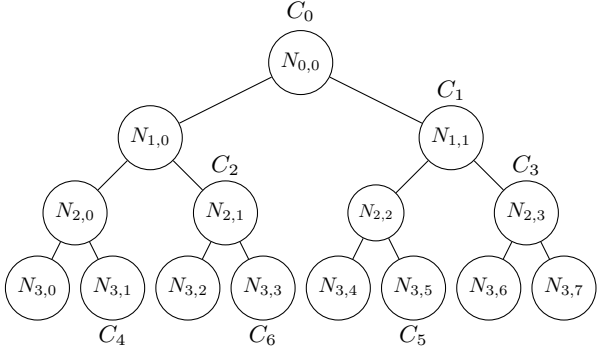
\begin{figure}[!t]
\centering
\begin{tikzpicture}[
level 1/.style={sibling distance = 4cm, level distance = 1cm},
level 2/.style={sibling distance = 2cm},
level 3/.style={sibling distance = 1cm},
label distance=-1mm,
every node/.style={circle, draw=black,thin, minimum size = 0.5cm},
emph/.style={edge from parent/.style={red,very thick,draw}},
norm/.style={edge from parent/.style={black,thin,draw}}
]
\node [label={[label distance=-1.5mm]450:{\small{$C_0$}}},scale=0.8]{$N_{0,0}$}
  child {node[label={[label distance=-2mm]450:{\small{$$}}},scale=0.8] {$N_{1,0}$}
    child {node[label={[label distance=-2mm]450:{\small{$$}}},scale=0.8] {$N_{2,0}$}
        child {node[label={[label distance=-3mm]270:{\small{$$}}},scale=0.8] {$N_{3,0}$}}
        child {node[label={[label distance=-2mm]270:{\small{$C_4$}}},scale=0.8] {$N_{3,1}$}}
    }
    child {node[label={[label distance=-2mm]450:{\small{$C_2$}}},scale=0.8] {$N_{2,1}$}
        child {node[label={[label distance=-3mm]270:{\small{$$}}},scale=0.8] {$N_{3,2}$}}
        child {node[label={[label distance=-2mm]270:{\small{$C_6$}}},scale=0.8] {$N_{3,3}$}}
    }
  }
  child {node[label={[label distance=-2mm]450:{\small{$C_1$}}},scale=0.8] {$N_{1,1}$}
    child {node[label={[label distance=-2mm]450:{\small{$$}}},scale=0.7] {$N_{2,2}$}
        child {node[label={[label distance=-3mm]270:{\small{$$}}},scale=0.8] {$N_{3,4}$}}
        child {node[label={[label distance=-2mm]270:{\small{$C_5$}}},scale=0.8] {$N_{3,5}$}}
    }
    child {node[label={[label distance=-2mm]450:{\small{$C_3$}}},scale=0.8] {$N_{2,3}$}
        child {node[label={[label distance=-3mm]270:{\small{$$}}},scale=0.8] {$N_{3,6}$}}
        child {node[label={[label distance=-3mm]270:{\small{$$}}},scale=0.8] {$N_{3,7}$}}
    }
  };
\end{tikzpicture}
\caption{Example of an initial task-to-core assignment for $c = 7$}
\label{virtual-top}
\end{figure}

In the initialization phase, for a binary search tree and the number of
cores a power of two ($c = 2^x$), one strategy would be to
generate all search-nodes at depth $x$ and assign one to each core. However, these requirements are too restrictive
and greatly complicate the implementation. Instead, we arrange the cores in a virtual tree-like topology
and force every core, except $C_0$, to request the first task from its parent (stored as a global variable)
in this virtual topology. $C_0$ is always assigned task $N_{0,0}$.
The {\sc GetParent} function is given in Figure \ref{get-parent}.
The intuition is that if we assume that cores join the computation in increasing order of rank,
$C_i$ must always request an initial task from $C_j$ where $j < i$ and there exists
no $C_k$ such that $k < i$ and $C_k$ has a heavier task than $C_j$.
Figure~\ref{virtual-top} shows
an example of an initial task-to-core assignment for $c = 7$.
The parent of $C_i$ corresponds to the first $C_j$ encountered on the path from the task assigned to
$C_i$ to the root.
When $C_4$ joins the topology, although all remaining cores ($C_0, \ldots, C_3$) have tasks of
equal weight, $C_4$ selects $C_0$ as a parent. This is due to the alternating behavior of
the {\sc GetParent} function. When $i$ is even, the parent of $C_i$ corresponds to $C_j$, where $j$ is
the smallest even integer such that $C_j$ has a task of maximum weight. The same holds for odd $i$, except
that $C_1$  must pick $C_0$ as a parent.
This approaches balances the number of cores exploring different sections of the search tree.

\begin{figure}[!t]
\centering
\begin{algorithmic}[1]
\Function{GetParent}{$r, c$}
\State $parent \gets 0$;
\For{$(i = 0;~ i < c;~ i++)$}
    \If{($2^i > r$)}
        \State \textbf{break};
    \EndIf
    \State $parent \gets r - 2^i$;
\EndFor
\State \textbf{return} $parent$;
\EndFunction
\end{algorithmic}
\begin{algorithmic}[1]
\Function{GetNextParent}{$r, c$}
\State $parent \gets (parent + 1) \mod c$;
\If{($parent = r$)}
    \State $parent \gets parent + 1$;
    \State $passes \gets passes + 1$
\EndIf
\State \textbf{return} $parent$;
\EndFunction
\end{algorithmic}
\caption{The {\sc GetParent} and {\sc GetNextParent} functions}
\label{get-parent}
\end{figure}

Once every core receives a response from its initial parent, the
initialization phase is complete.  After that, each core updates its
parent to $(r + 1)\mod c$.  During the search-phase, whenever a core
requires a new task, it will first attempt to request one from its
current parent. If the parent has no available tasks or is inactive,
the virtual topology is modified by the {\sc GetNextParent} function
(Figure \ref{get-parent}).  In the global variable $passes$, we keep
track of the number of times each core has unsuccessfully requested a
task from all participants. The termination protocol is fired by some
core $C_i$ whenever $passes > 2$.  $C_i$ goes from being active to
inactive and sends a status update message to inform the remaining
participants. Once all cores are inactive, the computation can safely
end.

The complete pseudocode for {\sc Parallel-RB} is given in Figure \ref{parallel-rb}.
The algorithm consists of two main procedures: {\sc Parallel-RB-Iterator} and  {\sc Parallel-RB-Solver}.
All communication must be non-blocking in the latter and blocking in the former.

\begin{figure}[!t]
\centering
\begin{algorithmic}[1]
\Procedure{Parallel-RB-Iterator}{$r, c$}
\State $init \gets true$;~$passes \gets 0$;
\State $parent \gets \Call{GetParent}{r, c}$;
\While{true}
    \If{($passes > 2$)}
        \State \Call{TerminationProtocol}{}();
    \EndIf
    \If{($r = 0$ \& $init$)}
        \State $init \leftarrow false$;
        \State \Call{Parallel-RB-Solver}{$N_{0,0}$};
    \Else
        \If{($init$)}
            \State $init \leftarrow false$;
            \State $idx \leftarrow \Call{RequestTaskIndex}{parent}$;
            \State $parent \gets (r + 1)\mod{c}$;
            \If{($idx \neq null$)}
                \State $N \leftarrow \Call{ConvertIndex}{idx}$;
                \State \Call{Parallel-RB-Solver}{$N$};
            \EndIf
        \EndIf
    \EndIf
    \State $parent \gets \Call{GetNextParent}{r, c}$;
    \State $idx \leftarrow \Call{RequestTaskIndex}{parent}$;
    \If{($idx \neq null$)}
        \State $N \leftarrow \Call{ConvertIndex}{idx}$;
        \State {\sc Parallel-RB-Solver}($N$);
    \EndIf
\EndWhile
\EndProcedure
\Statex
\Procedure{Parallel-RB-Solver}{$N_{d,p}$}
\If{($current\_idx[d] = -1$)}
  \State \textbf{terminate};
\EndIf
\State $current\_idx[d] \gets p$;
\If{(\Call{IsSolution}{$N_{d,p}$})}
  \State $best\_so\_far \gets N_{d,p}$;
  \State Broadcast the new solution;\Comment{Optional}
\EndIf
\If{(\Call{IsLeaf}{$N_{d,p}$})}
  \State Apply backtracking;\Comment{undo operations}
\EndIf
\If{(\Call{TaskRequestExists}{}())}
    \State $x \leftarrow \Call{GetHeaviestTaskIndex}{current\_idx}$;
    \State \Call{send}{$x$, $requester$};
\EndIf
\If{(\Call{BroadcastMessageExists}{}())}
    \State Read and perform necessary actions;
\EndIf
\State $p' \gets 0$;
\While{\Call{HasNextChild}{$N_{d,p}$}}
    \State $N_{d + 1, p'} \gets \Call{GetNextChild}{N_{d,p}}$;
    \State \Call{Parallel-RB-Solver}{$N_{d + 1, p'}$};
    \State $p' \gets p' + 1$;
\EndWhile
\EndProcedure
\end{algorithmic}
\caption{The {\sc Parallel-RB} algorithm}
\label{parallel-rb}
\end{figure}

\subsection{Arbitrary Branching Factor}\label{sec-not-binary}
For search trees of arbitrary branching factor, the index of $N_{d,p}$ needs to keep track of
both the unique root-to-node path as well as the number of unexplored siblings of $N_{d,p}$ (i.e. all the nodes at depth
$d$ and position greater than $p$). Therefore, we divide an index into two parts, $idx_1$ and $idx_2$.
We let $k^{th}(N_{d,p})$ denote the $k^{th}$ child of $N_{d,p}$
and $C(N_{d,p})$ the set of all children of $N_{d,p}$.
The following procedure assigns indices to every search-node in $T$:

\begin{itemize}
\item[(1)] The root of $T$ has $idx_1(N_{0,0}) = 1$ and $idx_2(N_{0,0}) = 0$
\item[(2)] For any node $N_{d,p}$ in $T$:
\begin{itemize}
\item[] $idx_1(k^{th}(N_{d,p})) = idx_1(N_{d,p}) \mid\mid (k - 1)$ and
\item[] $idx_2(k^{th}(N_{d,p})) = idx_2(N_{d,p}) \mid\mid (|C(N_{d,p})| - k)$
\end{itemize}
\end{itemize}

\begin{figure}[!t]
\centering
\begin{tikzpicture}[
level 1/.style={sibling distance = 2cm},
level 2/.style={sibling distance = 2cm},
level 3/.style={sibling distance = 2cm},
label distance=-1mm,
every node/.style={circle, draw = black, thin, minimum size = 0.5cm},
emph/.style={edge from parent/.style={red,very thick,draw}},
norm/.style={edge from parent/.style={black,thin,draw}}
]
\node [label={[label distance=-4mm]450:{\small{    \begin{tabular}{l}$1$\\$0$\end{tabular}    }}},scale=0.8]{$N_{0,0}$}
  child {node[label={[label distance=-4mm]450:{\small{    \begin{tabular}{l}$10$\\$02$\end{tabular}    }}},scale=0.8] {$N_{1,0}$}
    child {node[label={[label distance=-4mm]360:{\small{    \begin{tabular}{l}$100$\\$021$\end{tabular}    }}},scale=0.8] {$N_{2,0}$}
        child {node[label={[label distance=-4mm]270:{\small{    \begin{tabular}{l}$1000$\\$0212$\end{tabular}    }}},scale=0.8] {$N_{3,0}$}}
        child {node[label={[label distance=-4mm]270:{\small{    \begin{tabular}{l}$1001$\\$0211$\end{tabular}    }}},scale=0.8] {$N_{3,1}$}}
        child {node[label={[label distance=-4mm]270:{\small{    \begin{tabular}{l}$1002$\\$0210$\end{tabular}    }}},scale=0.8] {$N_{3,2}$}}
    }
    child {node[label={[label distance=-4mm]360:{\small{    \begin{tabular}{l}$101$\\$020$\end{tabular}    }}},scale=0.8] {$N_{2,1}$}}
  }
  child {node[label={[label distance=-4mm]440:{\small{    \begin{tabular}{l}$11$\\$01$\end{tabular}    }}},scale=0.8] {$N_{1,1}$}}
  child {node[label={[label distance=-4mm]450:{\small{    \begin{tabular}{l}$12$\\$00$\end{tabular}    }}},scale=0.8] {$N_{1,2}$}};
\end{tikzpicture}
\caption{Example of an indexed search tree with arbitrary branching factor}
\label{fig-tree2}
\end{figure}
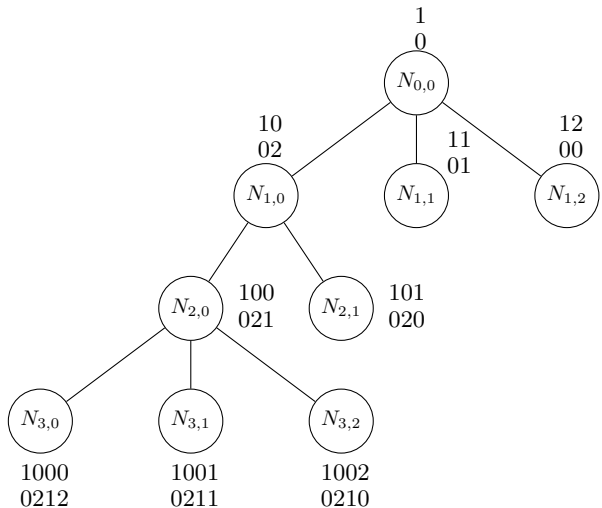

\noindent An example of an indexed search tree is given in Figure \ref{fig-tree2}.
Each node is assigned two identifiers: $idx_1$ (top) and $idx_2$ (bottom).
At the implementation level, the $current\_idx$ array is replaced by a $2 \times d$ array
that can be maintained after every recursive call in a similar fashion to {\sc Parallel-RB-Solver}
as long as each search-node $N_{d,p}$ is aware of $|C(N_{d,p})|$.
The first non-zero entry in $current\_idx[1]$ (the second row of the array), say $current\_idx[1][x]$,
indicates the depth of all tasks of heaviest weight.
Since there can be more than one unvisited node at this depth, we could
choose to send either all of them or just a subset $S$. In the first case, we can remember
delegated branches by simply setting $current\_idx[1][x]$ to $-1$.
For the second case, $current\_idx[1][x]$ is decremented by $|S|$.
Note that the choice of $S$ cannot be arbitrary. If
$C(N_{d,p}) = \{ N_{d + 1,0}, N_{d + 1,1}, \ldots, N_{d + 1,p_{max}}\}$,
$S$ must include $N_{d + 1,p_{max}}$, and for any $N_{d + 1,i} \in S$, it must be the case that
$N_{d + 1,j}$ is also in $S$ for all $j$ between $i$ and $p_{max}$.
The only modification required in
{\sc Parallel-RB-Solver} is to make sure that at search-node $N_{x,p}$,
{\sc GetNextChild} is executed only $current\_idx[1][x]$ times.

\section{Implementation}\label{implementation}
We tested our framework with parallel implementations of algorithms for the well-known
{\sc Vertex Cover} and {\sc Dominating Set} problems.

\vspace{8pt}
\begin{tabular}{ll}
\multicolumn{2}{l}{{\sc Vertex Cover}}\\
{\bf Input}: & A graph $G = (V, E)$\\
{\bf Question}: & Find a set $C \subseteq V$ such that $|C|$ is\\
& minimized and the graph induced by $V \setminus C$\\
& is edgeless
\end{tabular}
\vspace{8pt}

\vspace{8pt}
\begin{tabular}{ll}
\multicolumn{2}{l}{{\sc Dominating Set}}\\
{\bf Input}: & A graph $G = (V, E)$\\
{\bf Question}: & Find a set $D \subseteq V$ such that $|D|$ is\\
& minimized and every vertex in $G$ is either\\
& in $D$ or is adjacent to a vertex in $D$
\end{tabular}
\vspace{8pt}

Both problems have received considerable attention in the areas of
exact and fixed parameter algorithms because of their close relations
to many other problems in different application domains \cite{ALSS06}.
The sequential algorithm for the parameterized version of {\sc Vertex
  Cover} having the fastest known worst-case behavior runs in
${\mathcal{O}}(kn + 1.2738^k)$ time~\cite{CKX10}. We converted this to
an optimized version by introducing simple modifications and
excluding complex processing rules that require heavy maintenance
operations.  For {\sc Dominating Set}, we implemented the algorithm of
\cite{FGK05} where the problem is solved by a reduction to {\sc
  Minimum Set Cover}.  We used the hybrid graph
data-structure~\cite{ALMN10} which was specifically designed for
recursive backtracking algorithms that combines the advantages of the
two classical adjacency-list and adjacency-matrix representations of
graphs with very efficient implicit backtracking operations.

Our input consists of a graph $G = (V, E)$ where $|V| = n$, $|E| = m$, and each vertex is given
an identifier between $0$ and $n - 1$.
The search tree for each algorithm is binary and the actual implementations
closely follow the {\sc Parallel-RB} algorithm. At every search-node, a vertex $v$
of highest degree is selected deterministically.
Vertex selection has to be deterministic to meet the requirements of the framework.
To break ties when multiple vertices have the same degree, we always
pick the vertex with the smallest identifier.
For {\sc Vertex Cover}, the left branch adds $v$ to the solution and the right branch adds
all the neighbors of $v$ to the solution. For {\sc Dominating Set}, the left branch is identical
but the right branch forces $v$ to be out of any solution.
The {\sc ConvertIndex} function is straightforward as the added, deleted, or discarded vertices
can be retraced by iterating through the index and applying
any appropriate reduction rules along the way.
Every time a smaller solution is found, the size is broadcasted to
all participants to avoid exploring branches that cannot lead to any improvements.

\section{Experimental Results}\label{exp-results}
All our code relies on the Message Passing Interface (MPI) \cite{DW94}
and uses the standard C language with no other dependencies.
Computations were performed on the BGQ supercomputer 
at the SciNet HPC Consortium\footnote{SciNet is funded by the Canada 
Foundation for Innovation under the auspices of Compute
Canada; the Government of Ontario; Ontario Research Fund - Research Excellence; and the University
of Toronto. \cite{LOKEN10}}. The BGQ production system is a $3^{rd}$ generation Blue Gene IBM supercomputer
built around a system-on-a-chip compute node. There are 2,048 compute nodes each
having a 16 core 1.6GHz PowerPC based CPU with 16GB of RAM. When running jobs on 32,768 cores,
each core is allocated 1GB of RAM. Each core also has four ``hardware threads'' which can keep the different
parts of each core busy at the same time. It is therefore possible to run jobs
on $65,536$ and $131,072$ cores at the cost of reducing available RAM per core to 500MB and 250MB, respectively.
We could run experiments using this many cores only when the input
graph was relatively small and, due to the fact
that multiple cores were forced to share (memory and CPU) resources, we
noticed a slight decrease in performance.

The {\sc Parallel-Vertex-Cover} algorithm was tested on four input graphs.
\begin{itemize}
\item[-] p\_hat700-1.clq: is a graph on $700$ vertices and $234,234$ edges with a minimum vertex cover of size $635$
\item[-] p\_hat1000-2.clq is a graph on $1,000$ vertices and $244,799$ edges with a minimum vertex cover of size $946$
\item[-] frb30-15-1.mis is a graph on $450$ vertices and $17,827$ edges with a minimum vertex cover of size $420$
\item[-] 60-cell is a graph on $300$ vertices and $600$ edges with a minimum vertex cover of size $190$
\end{itemize}
The first two instances were obtained from the classical
Center for Discrete Mathematics and Theoretical Computer Science (DIMACS) benchmark suite
(\url{http://dimacs.rutgers.edu/Challenges/}).
The frb30-15-1.mis graph is a notoriously hard instance for which the exact size of
a solution was only known so far from a theoretical perspective.
To the best of our knowledge, this paper is the first to experimentally solve it; more information on this instance can be found in \cite{XBHL05}.
Lastly, the 60-cell graph is a 4-regular graph (every vertex has exactly $4$ neighbors) with applications
in chemistry \cite{DDMSWWFDD13}. Prior to this work, we solved the 60-cell using a serial algorithm
which ran for almost a full week \cite{ALMN10}. The high regularity of the graph makes it very hard
to apply any pruning rules resulting in an almost exhaustive enumeration of all feasible solutions.
For the {\sc Parallel-Dominating-Set} algorithm we generated two random instances 201x1500.ds and 251x6000.ds
where $n$x$m$.ds denotes a graph on $n$ vertices and $m$ edges.
Both instances could not be solved by our serial algorithm when limited to 24 hours.

\begin{table}[!t]
\renewcommand{\arraystretch}{1.3}
\caption{{\sc Parallel-Vertex-Cover} Statistics}
\label{table-vc-results}
\centering
\begin{tabular}{|c|c|c|c|c|}
\hline
\textbf{Graph} & $\mathbf{|C|}$ & \textbf{Time} & $\mathbf{T_S}$ & $\mathbf{T_R}$\\
\hline
p\_hat700-1.clq & 16 & 19.5hrs & 2,876 & 2,910\\
\hline
p\_hat700-1.clq & 32 & 9.8hrs & 2,502 & 2,567\\
\hline
p\_hat700-1.clq & 64 & 4.9hrs & 3,398 & 3,518\\
\hline
p\_hat700-1.clq & 128 & 2.5hrs & 4,928 & 5,196\\
\hline
p\_hat700-1.clq & 256 & 1.3hrs & 4,578 & 5,153\\
\hline
p\_hat700-1.clq & 512 & 38min & 4,354 & 5,451\\
\hline
p\_hat700-1.clq & 1,024 & 18.9min & 4,052 & 6,391\\
\hline
p\_hat700-1.clq & 2,048 & 9.89min & 3,781 & 8,117\\
\hline
p\_hat700-1.clq & 4,096 & 5.39min & 3,665 & 11,978\\
\hline
p\_hat700-1.clq & 8,192 & 2.9min & 2,714 & 19,183\\
\hline
p\_hat700-1.clq & 16,384 & 1.7min & 1,342 & 32,883\\
\hline
p\_hat1000-2.clq & 64 & 23.6min & 3,664 & 3,799\\
\hline
p\_hat1000-2.clq & 128 & 12.5min & 2,651 & 2,912\\
\hline
p\_hat1000-2.clq & 256 & 6.5min & 1,623 & 1,956\\
\hline
p\_hat1000-2.clq & 512 & 3.7min & 1,235 & 1,872\\
\hline
p\_hat1000-2.clq & 1,024 & 2.1min & 866 & 2,142\\
\hline
p\_hat1000-2.clq & 2,048 & 1.2min & 610 & 3,120\\
\hline
frb30-15-1.mis & 1,024 & 14.2hrs & 13,580 & 15,968\\
\hline
frb30-15-1.mis & 2,048 & 7.2hrs & 21,899 & 26,597\\
\hline
frb30-15-1.mis & 4,096 & 3.6hrs & 28,740 & 37,733\\
\hline
frb30-15-1.mis & 8,192 & 1.9hrs & 29,110 & 45,685\\
\hline
frb30-15-1.mis & 16,384 & 55.1min & 28,707 & 59,978\\
\hline
frb30-15-1.mis & 32,768 & 28.8min & 30,008 & 96,438\\
\hline
frb30-15-1.mis & 65,536 & 16.8min & 25,359 & 158,371\\
\hline
frb30-15-1.mis & 131,072 & 11.1min & 19,419 & 312,430\\
\hline
60-cell & 128 & 14.3hrs & 19 & 26\\
\hline
60-cell & 256 & 7.3hrs & 23 & 23\\
\hline
60-cell & 512 & 3.7hrs & 1,091 & 1,388\\
\hline
60-cell & 1,024 & 45.1min & 1,397 & 1,940\\
\hline
60-cell & 2,048 & 11.3min & 1,331 & 2,430\\
\hline
60-cell & 4,096 & 2.8min & 949 & 3,094\\
\hline
\end{tabular}
\end{table}

\begin{table}[!t]
\renewcommand{\arraystretch}{1.3}
\caption{{\sc Parallel-Dominating-Set} Statistics}
\label{table-ds-results}
\centering
\begin{tabular}{|c|c|c|c|c|}
\hline
\textbf{Graph} & $\mathbf{|C|}$ & \textbf{Time} & $\mathbf{T_S}$ & $\mathbf{T_R}$\\
\hline
201x1500.ds & 512 & 18.1hrs & 8,231 & 9,642\\
\hline
201x1500.ds & 1,024 & 9.2hrs & 10,315 & 12,611\\
\hline
201x1500.ds & 2,048 & 4.5hrs & 11,566 & 16,118\\
\hline
201x1500.ds & 4,096 & 2.3hrs & 14,070 & 23,413\\
\hline
201x1500.ds & 8,192 & 1.2hrs & 13,243 & 33,680\\
\hline
201x1500.ds & 16,384 & 36.2min & 10,295 & 41,795\\
\hline
201x1500.ds & 32,768 & 19.2min & 6,925 & 72,719\\
\hline
201x1500.ds & 65,536 & 11.8min & 4,221 & 109,346\\
\hline
251x6000.ds & 256 & 8.9hrs & 3,313 & 4,573\\
\hline
251x6000.ds & 512 & 4.7hrs & 3,865 & 4,985\\
\hline
251x6000.ds & 1,024 & 2.4hrs & 2,842 & 5,306\\
\hline
251x6000.ds & 2,048 & 1.2hrs & 1,528 & 5,396\\
\hline
251x6000.ds & 4,096 & 36.4min & 2,037 & 9,714\\
\hline
251x6000.ds & 8,192 & 18.7min & 1,445 & 10,497\\
\hline
251x6000.ds & 16,384 & 10.1min & 1,132 & 12,310\\
\hline
251x6000.ds & 32,768 & 5.5min & 934 & 13,982\\
\hline
\end{tabular}
\end{table}

All of our experiments were limited by the system to a maximum of 24 hours per job.
To evaluate the performance of our communication model and dynamic load balancing strategy,
we collected two statistics from each run: $T_S$ and $T_R$.
$T_S$ denotes the average number of tasks received (and hence solved) by each core
while $T_R$ denotes the average number of tasks requested by each core.
In Table \ref{table-vc-results}, we give the running
times of the {\sc Parallel-Vertex-Cover}
algorithm for every instance while varying the total number
of cores, $|C|$, from $2$ to $131,072$ (we only ran experiments
on 65,536 or 131,072 cores when the graph was small
enough to fit in memory or when the running time exceeded 10 minutes on 32,768 cores).

\begin{figure}[!t]
\centering
\includegraphics[width=3.5in]{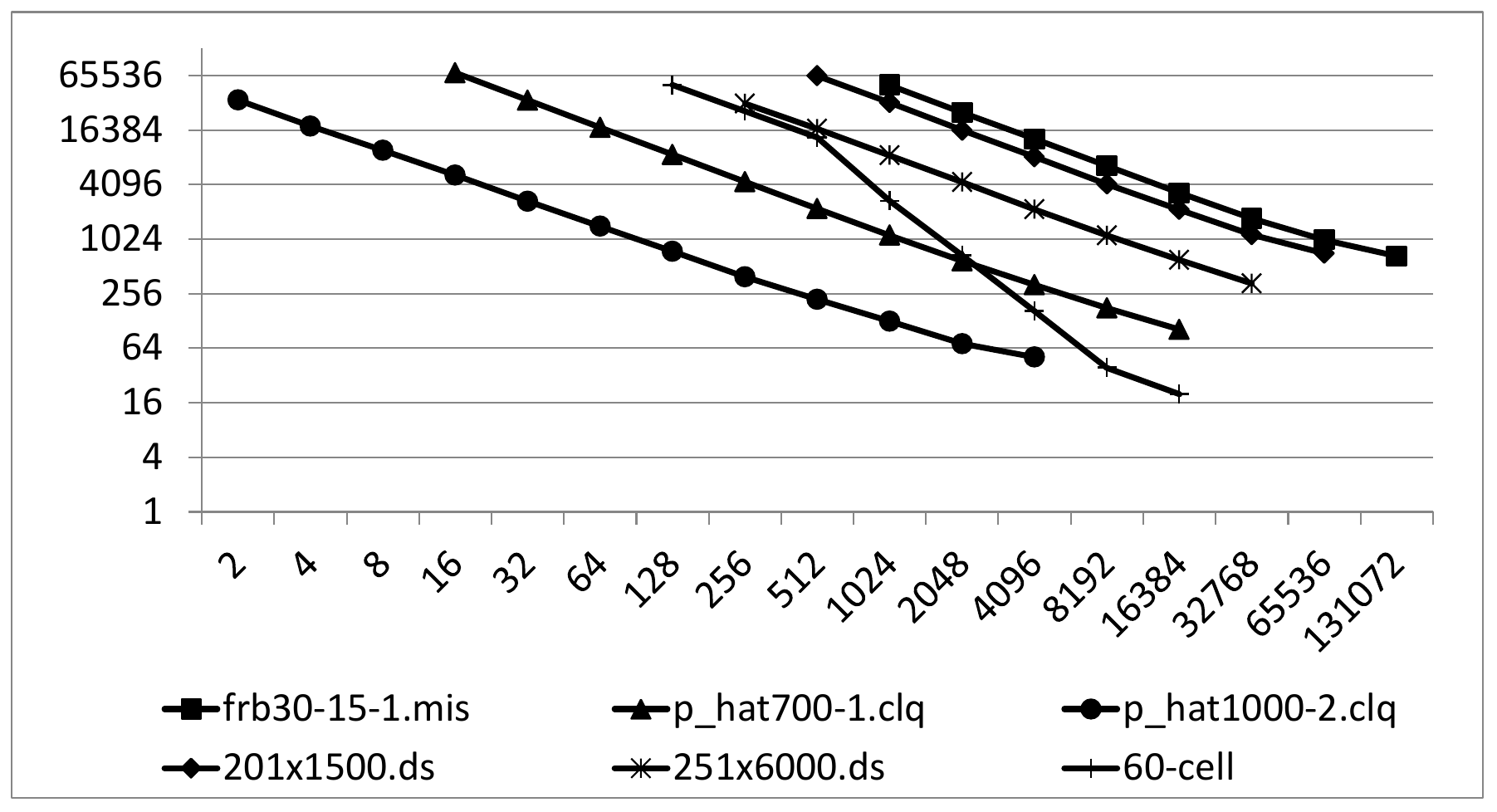}
\caption{The logarithm (base 2) of running times in seconds (y-axis) vs. number of cores (x-axis)}
\label{fig-results-chart1}
\end{figure}

\begin{figure}[!t]
\centering
\includegraphics[width=3.5in]{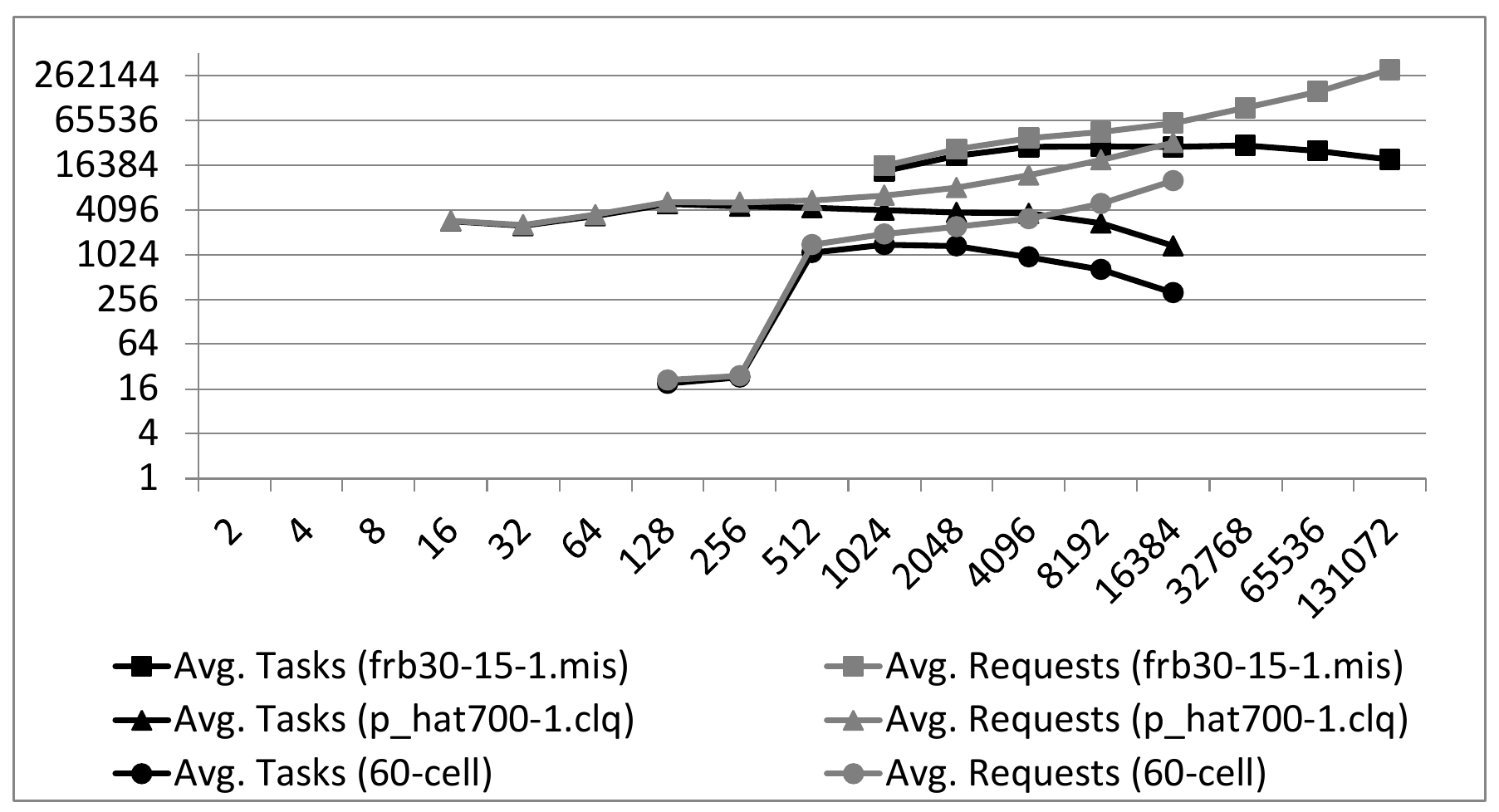}
\caption{The logarithm (base 2) of the average number of message transmissions (y-axis) vs. number of cores (x-axis)
($T_S$ shown in black and $T_R$ shown in gray)}
\label{fig-results-chart2}
\end{figure}

The values of $T_S$ and $T_R$ are also provided.
Similar results for the {\sc Parallel-Dominating-Set} algorithm are given
in Table \ref{table-ds-results}. Due to space limitations, we omit
some of the entries in the tables
and show the overall behaviors in the chart of Figure \ref{fig-results-chart1}.
In almost all cases, the algorithms achieve near linear (super-linear for the 60-cell graph)
speedup on at least 32,768 cores.
We were not able to gather enough experimental data to characterize 
the behavior of the framework on 65,536 and 131,072 cores but results on 201x1500.ds 
and frb30-15-1.mis suggest a 10 percent decrease in performance.
As noted earlier, cores were forced to share resources whenever $|C|$ 
was greater than 32,768. Further work is needed to determine whether 
this can explain the performance decrease or if other factors came into play.
We also note that harder instances are required to fairly test 
scalability on a larger number of cores since the ones we considered were all solved
in just a few minutes using at most 32,768 cores.
In Figure \ref{fig-results-chart2}, we plot the different values of $T_S$ and $T_R$ for
a representative subset of our experiments. This chart reveals the inherent difficulty of
dynamic load balancing. As $|C|$ increases, the gap between $T_S$ and $T_R$ grows larger and larger.
Any efficient dynamic load-balancing strategy has to control the growth of this gap (e.g. keep it linear)
for a chance to achieve unbounded scalability.
The largest gap we obtained was on the frb30-15-1.mis instance using 131,072 cores. The gap was
approximately 300,000. Given the number of cores
and the amount of time ($>$ 10 minutes) spent on the computation, the number suggests that each core
requested an average of 2.5 tasks from every other core.
One possible improvement which we are currently investigating is to modify our virtual topology
to a graph-like structure of bounded degree. The framework currently assumes a fully connected topology
after initialization (i.e. any two cores can communicate) which explains the correlation
between $|C|$ and $|T_S - T_R|$. By bounding the degrees in the virtual topology, we hope to make
this gap weakly dependent on $|C|$.

\section{Conclusions and Future Work}
Combining indexed search-trees with a decentralized communication model, we have showed
how any serial recursive backtracking algorithm, with some ordered branching,
can be modified to run in parallel.
Some of the key advantages of our framework are:
\begin{itemize}
\item[-] The migration from serial to parallel entails very little additional coding.
Implementing each of our parallel algorithms took less than two days and
around 300 extra lines of code.
\item[-] It completely eliminates the need for task-buffers and the overhead they introduce (Section \ref{sec-buffers}).
\item[-] The input of both the serial and parallel implementations are identical.
Running the parallel algorithms requires no additional input from the user (assuming
every core has access to $r$ and $c$). Most parallel algorithms in the literature require some parameters
such as task-buffer size. Selecting the best parameters could vary depending on the instance being solved.
\item[-] Experimental results have showed that our implicit load-balancing strategy,
joined with the concise task-encoding scheme, can achieve linear (sometimes super-linear) speedup with
scalability on at least $32,768$ cores. We hope to test our framework on a larger system in the near future to
determine the maximum number of cores it can support.
\item[-] Although not typical of parallel algorithms, when using the indexing method and the {\sc ConvertIndex} function,
it becomes reasonably straightforward to support join-leave (i.e. cores leaving the computation after solving a fixed number of tasks)
or checkpointing capabilities (i.e. by forcing every core to write its $current\_idx$ to some file).
\end{itemize}

We believe there is still plenty of room for improving the framework at the risk of losing some of its
simplicity. One area we have started examining is the virtual topology. A ``smarter'' topology could
further reduce the communication overhead (e.g. the gap between $T_S$ and $T_R$) and
increase the overall performance. Another candidate is the {\sc GetNextParent} function which
can be modified to probe a fixed number of cores before selecting which to ``help'' next.
Finally, we intend to investigate the possibility of developing the
framework into a library, similar to \cite{RLS04} and \cite{SZJK11},
which will provide users with built-in functions
for parallelizing recursive backtracking algorithms.

\section*{Acknowledgment}
The authors would like to thank Chris Loken and the SciNet
team for providing access to the BGQ production system
and for their support throughout the experiments.



\bibliographystyle{IEEEtran}
\bibliography{references}

\begin{thebibliography}{10}
\providecommand{\url}[1]{#1}
\csname url@samestyle\endcsname
\providecommand{\newblock}{\relax}
\providecommand{\bibinfo}[2]{#2}
\providecommand{\BIBentrySTDinterwordspacing}{\spaceskip=0pt\relax}
\providecommand{\BIBentryALTinterwordstretchfactor}{4}
\providecommand{\BIBentryALTinterwordspacing}{\spaceskip=\fontdimen2\font plus
\BIBentryALTinterwordstretchfactor\fontdimen3\font minus
  \fontdimen4\font\relax}
\providecommand{\BIBforeignlanguage}[2]{{%
\expandafter\ifx\csname l@#1\endcsname\relax
\typeout{** WARNING: IEEEtran.bst: No hyphenation pattern has been}%
\typeout{** loaded for the language `#1'. Using the pattern for}%
\typeout{** the default language instead.}%
\else
\language=\csname l@#1\endcsname
\fi
#2}}
\providecommand{\BIBdecl}{\relax}
\BIBdecl

\bibitem{CKJ01}
J.~Chen, I.~A. Kanj, and W.~Jia, ``Vertex cover: further observations and
  further improvements,'' \emph{Journal of Algorithms}, vol.~41, no.~2, pp.
  280--301, 2001.

\bibitem{FKW05}
F.~V. Fomin, D.~Kratsch, and G.~J. Woeginger, ``Exact (exponential) algorithms
  for the dominating set problem,'' in \emph{Graph-Theoretic Concepts in
  Computer Science}, 2005, vol. 3353, pp. 245--256.

\bibitem{CKX10}
J.~Chen, I.~A. Kanj, and G.~Xia, ``Improved upper bounds for vertex cover,''
  \emph{Theor. Comput. Sci.}, vol. 411, no. 40-42, pp. 3736--3756, Sep. 2010.

\bibitem{FGK05}
F.~V. Fomin, F.~Grandoni, and D.~Kratsch, ``Measure and conquer: Domination – a
  case study,'' in \emph{Automata, Languages and Programming}, 2005, vol. 3580,
  pp. 191--203.

\bibitem{RND09}
J.~M.~M. Rooij, J.~Nederlof, and T.~C. Dijk, ``Inclusion/exclusion meets
  measure and conquer,'' in \emph{Algorithms - ESA 2009}, 2009, vol. 5757, pp.
  554--565.

\bibitem{W03}
G.~J. Woeginger, ``Exact algorithms for {NP}-hard problems: a survey,'' in
  \emph{Combinatorial optimization - Eureka, you shrink!}, 2003, pp. 185--207.

\bibitem{DF97}
R.~G. Downey and M.~R. Fellows, \emph{Parameterized complexity}.\hskip 1em plus
  0.5em minus 0.4em\relax New York: Springer-Verlag, 1997.

\bibitem{DG08}
J.~Dean and S.~Ghemawat, ``{M}ap{R}educe: simplified data processing on large
  clusters,'' \emph{Commun. ACM}, vol.~51, no.~1, pp. 107--113, Jan. 2008.

\bibitem{DB07}
G.~D. Fatta and M.~R. Berthold, ``Decentralized load balancing for highly
  irregular search problems,'' \emph{Microprocess. Microsyst.}, vol.~31, no.~4,
  pp. 273--281, Jun. 2007.

\bibitem{WERL11}
D.~P. Weerapurage, J.~D. Eblen, G.~L. Rogers, Jr., and M.~A. Langston,
  ``Parallel vertex cover: a case study in dynamic load balancing,'' in
  \emph{Proceedings of the Ninth Australasian Symposium on Parallel and
  Distributed Computing}, vol. 118, 2011, pp. 25--32.

\bibitem{AM12}
F.~N. Abu-Khzam and A.~E. Mouawad, ``A decentralized load balancing approach
  for parallel search-tree optimization,'' in \emph{Proceedings of the 2012
  13th International Conference on Parallel and Distributed Computing,
  Applications and Technologies}, 2012, pp. 173--178.

\bibitem{K88}
L.~V. Kale, ``Comparing the performance of two dynamic load distribution
  methods,'' in \emph{Proceedings of the 1988 International Conference on
  Parallel Processing}, 1988, pp. 8--11.

\bibitem{SZJK11}
Y.~Sun, G.~Zheng, P.~Jetley, and L.~V. Kale, ``An adaptive framework for
  large-scale state space search,'' in \emph{Proceedings of the 2011 IEEE
  International Symposium on Parallel and Distributed Processing Workshops and
  PhD Forum}, 2011, pp. 1798--1805.

\bibitem{RLS04}
T.~K. Ralphs, L.~L\'{a}danyi, and M.~J. Saltzman, ``A library hierarchy for
  implementing scalable parallel search algorithms,'' \emph{J. Supercomput.},
  vol.~28, no.~2, pp. 215--234, May 2004.

\bibitem{ARAS08}
F.~N. Abu-Khzam, M.~A. Rizk, D.~A. Abdallah, and N.~F. Samatova, ``The buffered
  work-pool approach for search-tree based optimization algorithms,'' in
  \emph{Proceedings of the 7th international conference on Parallel processing
  and applied mathematics}, 2008, pp. 170--179.

\bibitem{DDMSWWFDD13}
S.~Debroni, E.~Delisle, W.~Myrvold, A.~Sethi, J.~Whitney, J.~Woodcock, P.~W.
  Fowler, B.~de~La~Vaissiere, and M.~Deza, ``Maximum independent sets of the
  120-cell and other regular polyhedral,'' \emph{Ars Mathematica
  Contemporanea}, vol.~6, no.~2, pp. 197--210, 2013.

\bibitem{ALMN10}
F.~N. Abu-Khzam, M.~A. Langston, A.~E. Mouawad, and C.~P. Nolan, ``A hybrid
  graph representation for recursive backtracking algorithms,'' in
  \emph{Proceedings of the 4th international conference on Frontiers in
  algorithmics}, 2010, pp. 136--147.

\bibitem{FM87}
R.~Finkel and U.~Manber, ``{DIB} - a distributed implementation of
  backtracking,'' \emph{ACM Trans. Program. Lang. Syst.}, vol.~9, no.~2, pp.
  235--256, Mar. 1987.

\bibitem{KGV94}
V.~Kumar, A.~Y. Grama, and N.~R. Vempaty, ``Scalable load balancing techniques
  for parallel computers,'' \emph{J. Parallel Distrib. Comput.}, vol.~22,
  no.~1, pp. 60--79, Jul. 1994.

\bibitem{ALSS06}
F.~N. Abu-Khzam, M.~A. Langston, P.~Shanbhag, and C.~T. Symons, ``Scalable
  parallel algorithms for {FPT} problems,'' \emph{Algorithmica}, vol.~45,
  no.~3, pp. 269--284, Jul. 2006.

\bibitem{DW94}
J.~J. Dongarra and D.~W. Walker, ``{MPI}: A message-passing interface
  standard,'' \emph{International Journal of Supercomputing Applications},
  vol.~8, no. 3/4, pp. 159--416, 1994.

\bibitem{LOKEN10}
C.~Loken, D.~Gruner, L.~Groer, R.~Peltier, N.~Bunn, M.~Craig, T.~Henriques,
  J.~Dempsey, C.-H. Yu, J.~Chen, L.~J. Dursi, J.~Chong, S.~Northrup, J.~Pinto,
  N.~Knecht, and R.~V. Zon, ``Scinet: Lessons learned from building a
  power-efficient top-20 system and data centre,'' \emph{Journal of Physics:
  Conference Series}, vol. 256, no.~1, p. 012026, 2010.

\bibitem{XBHL05}
K.~Xu, F.~Boussemart, F.~Hemery, and C.~Lecoutre, ``A simple model to generate
  hard satisfiable instances,'' in \emph{Proceedings of the 19th international
  joint conference on Artificial intelligence}, 2005, pp. 337--342.

\end{thebibliography}
%

\end{document}